 \documentclass[preprint,aps,showpacs]{revtex4}
 
\begin{document}
 \newcommand{\Qed}{\rule{2.5mm}{3mm}}
 \newcommand{\balpha}{\mbox{\boldmath {$\alpha$}}}
 \def\Tr{{\rm Tr}}
 \def\(#1)#2{{\stackrel{#2}{(#1)}}}
 \def\[#1]#2{{\stackrel{#2}{[#1]}}}
 \def\A{{\cal A}}
  \def\B{{\cal B}}
 \def\Sb#1{_{\lower 1.5pt \hbox{$\scriptstyle#1$}}}
 \draft 
\title{Involution requirement on a boundary makes massless  fermions compactified on a finite 
flat disk mass protected}
\author{ N.S. Manko\v c Bor\v stnik}
\address{ Department of Physics, FMF, University of
Ljubljana, Jadranska 19,Ljubljana, 1000}
\author{ H. B. Nielsen}
\address{Department of Physics, Niels Bohr Institute,
Blegdamsvej 17,\\
Copenhagen, DK-2100}

\begin{abstract} 
The genuine Kaluza-Klein-like theories---with no fields in addition to gravity---have difficulties 
with the existence of massless spinors after the compactification 
of some space dimensions \cite{witten}. We proposed in ref.\ \cite{hnkk06} such a  
boundary condition for spinors in $1+5$ compactified on a flat disk that ensures  
masslessness of  spinors in $d=1+3$ as well as their   
chiral coupling to the corresponding background gauge gravitational field.  
In this paper we study the same toy model, 
looking this time 
for an involution  which  
transforms a space of solutions of Weyl equations in $d=1+5$ 
from the outside of the flat disk in $x^5$ and $x^6$ 
into its inside, 
allowing massless spinor of only one handedness---and accordingly assures mass protection---
and of one charge---$1/2$---and  
infinitely many massive spinors of the same charge. 
We reformulate the 
operator of momentum so that it is Hermitean on the vector space of spinor states obeying  
the involution boundary condition.
\end{abstract}


\pacs{Unifying theories(12.10.-g,04.60.Kz,11.15.-Ex), 
Kaluza-Klein-like theories(11.25.Mj,11.40.-q), 
mass protection mechanism(11.30.Rd), generalized Hermitean 
operators for momentum(03.65.-w), higher dimensional spaces(04.50.+h) }

\maketitle

\date{today}

\section{Introduction}
\label{introduction}

The major problem of the compactification procedure in all Kaluza-Klein-like theories with 
only gravity and no additional gauge fields is how to ensure that massless spinors 
be mass protected after the compactification. Namely, even if we start with 
only one Weyl spinor in some even dimensional space of  $d=2$ 
modulo $4$ dimensions (i.e.\ in $d=2(2n+1),$ $n=0,1,2,\cdots$) so that there appear no Majorana 
mass if no conserved 
charges exist and families are allowed, as we have proven in 
ref.\ \cite{hnm06}, and accordingly with the mass protection from the very beginning, 
a compactification of $m$ dimensions gives rise to a spinor of one handedness in $d$  
with both handedness in $d-m$ and is accordingly  not mass protected any longer. 

And  besides, since the spin (or the total conserved angular momentum) 
in the compactified part of space will    
in $d-m$ space manifest as a charge of a positive and a negative value  
and since in the second quantization procedure antiparticles 
of opposite charges appear anyhow, doubling the number of massless 
spinors when coming from 
$d(=2(2n+1))$-dimensional space down to $d=4$ and after 
the second quantized procedure is not in agreement  
with what we observe. Accordingly there must be some requirements, 
some boundary conditions, which 
ensure in a compactification procedure that only spinors of one handedness survive, if 
Kaluza-Klein-like theories have some meaning. However, the idea of Kaluza and Klein of 
having only gravity as a gauge field seems  too beautiful not to have the realization in Nature. 

One of us\cite{norma92,norma93,norma94,norma95,Portoroz03,pikanorma06} has for long tried to 
unify the spin and all the charges to only the spin, so that spinors would in 
$d\ge 4$ carry nothing but 
a spin and interact accordingly with only the gauge fields of the Poincar\' e group, that is with 
vielbeins $f^{\alpha}{\!}_{a}$ \footnote{$f^{\alpha}{}_{a}$ are inverted vielbeins to 
$e^{a}{}_{\alpha}$ with the properties $e^a{}_{\alpha} f^{\alpha}{\!}_b = \delta^a{\!}_b,\; 
e^a{\!}_{\alpha} f^{\beta}{\!}_a = \delta^{\beta}_{\alpha} $. 
Latin indices  
$a,b,..,m,n,..,s,t,..$ denote a tangent space (a flat index),
while Greek indices $\alpha, \beta,..,\mu, \nu,.. \sigma,\tau ..$ denote an Einstein 
index (a curved index). Letters  from the beginning of both the alphabets
indicate a general index ($a,b,c,..$   and $\alpha, \beta, \gamma,.. $ ), 
from the middle of both the alphabets   
the observed dimensions $0,1,2,3$ ($m,n,..$ and $\mu,\nu,..$), indices from 
the bottom of the alphabets
indicate the compactified dimensions ($s,t,..$ and $\sigma,\tau,..$). 
We assume the signature $\eta^{ab} =
diag\{1,-1,-1,\cdots,-1\}$.
} and   
spin connections $\omega_{ab\alpha}$, which are the gauge fields of the Poincar\'e group. 

In this paper  we take (as we did in ref.\ \cite{hnkk06}) the covariant momentum of a spinor, 
when applied on a spinor function $\psi$, to be 
\begin{eqnarray}
p_{0 a} &=& f^{\alpha}{\!}_{a}p_{0 \alpha}, \quad p_{0 \alpha} \psi = p_{ \alpha} - \frac{1}{2} S^{cd} 
\omega_{cd \alpha}. 
\label{covp}
\end{eqnarray}
The corresponding Lagrange density ${\cal L}$  for   a Weyl spinor has the form
${\cal L} = E \frac{1}{2} [(\psi^{\dagger}\gamma^0 \gamma^a p_{0a} \psi) + 
(\psi^{\dagger} \gamma^0\gamma^a p_{0 a}
\psi)^{\dagger}]$ and leads to
\begin{eqnarray}
{\cal L} &=& E\psi^{\dagger}\gamma^0 \gamma^a   ( p_{a} - \frac{1}{2} S^{cd}  \omega_{cda})\psi,
\label{weylL}
\end{eqnarray}
with $ E = \det(e^a{\!}_{\alpha}). $

The authors of this work have tried to find a way out of this ``Witten's no go theorem'' 
 for a toy model of 
$M^{(1+3)} \times$ a flat finite disk in $(1+5)$-dimensional space \cite{hnkk06} by postulating a 
particular boundary condition, which allows a spinor to carry  only one handedness 
after the compactification. 
Massless spinors then chirally couple to the corresponding background gauge gravitational field, 
which solves equations of motion for a free field, linear in the Riemann curvature, while 
the current through the wall for the massless and all the massive solutions is equal to zero. 

In ref.\ \cite{hnkk06} the boundary condition was written in a covariant way  as 
\begin{eqnarray}
\hat{\cal{R}}\psi|_{\rm wall} &=& 0,\quad 
\hat{\cal{R}} = \frac{1}{2}(1-i n^{(\rho)}{\!\!}_{a}\, n^{(\phi)}{\!\!}_{b}\, 
\gamma^a \gamma^b ),\quad \hat{\cal{R}}^2 = \hat{\cal{R}}
\label{diskboundary}
\end{eqnarray}
with $n^{(\rho)}=(0,0,0,0,\cos \phi, \sin \phi),\; n^{(\phi)}= 
(0,0,0,0,-\sin \phi, \cos \phi)$, which  
are the two unit vectors 
perpendicular and tangential to the boundary of the disk at $\rho_0$, respectively. 
The projector $\hat{\cal{R}}$ can for the above choice of the two vectors 
$n^{(\rho)}$ and $n^{(\phi)}$ be written as  
\begin{eqnarray}
\hat{\cal{R}}   &=& {\stackrel{56}{[-]} }= \frac{1}{2} (1-i\gamma^5\gamma^6).
\label{prodiskboundary}
\end{eqnarray}
The reader can find more about the Clifford algebra objects 
$\stackrel{ab}{(\pm)}, \stackrel{ab}{[\pm]}$ in Appendix \ref{appendixtechnique}. 

The boundary condition requires that only  massless states (determined by  
Eq.(\ref{weylL})) of one (let us say right) handedness 
with respect to the compactified disk  are allowed. Accordingly 
massless states of only one handedness are allowed in $d=1+3$.

In this paper: \\
i) {\em  We reformulate the  boundary condition as an involution}, which 
transforms the solutions of the equations of motion (or their superpositions) 
from outside the boundary of the disk 
into its inside. We do this with the intention that the limitation of 
$M^2$ on a finite disk  have 
a natural explanation, originated in a symmetry relation,    
allowing only {\em one massless state with the charge determined by the spin} 
in $x^5, x^6$  and 
infinitely many massive states with the same charge---so that to each mass only 
one state corresponds. \\
ii) {\em We redefine the definition of the momentum $p^s$ so that it becomes 
Hermitean on the vector space of states fulfilling the involution boundary conditions} 
and we comment on the orthogonality relations of these states.

We  make use of the technique presented in refs.\ \cite{holgernorma2002,holgernorma2003}
when writing the equations of motion and their solutions. We briefly repeat this technique in  
Appendix \ref{appendixtechnique}. It turns out 
that all the  derivations and discussions appear to be very transparent when using this technique.

\section{Equations of motion and solutions}
\label{equations}

We assume that the two dimensional space of coordinates $x^5$ and $x^6$ is a  
Euclidean plane $M^{(2)}$ (with no gravity)   
$f^{\sigma}{\!}_{s} = \delta^{\sigma}{\!}_{s},\; \omega_{56 s} =0$ 
and  with the rotational symmetry around an origin. 

Wave functions  describing spinors in $(1+5)$-dimensional space demonstrating  
$M^{(1+3)}$ $\times M^{(2)}$ symmetry  are 
required to obey the equations of motion
\begin{eqnarray}
\gamma^0 \gamma^a p_a \psi^{(6)} =0, \quad a=m,s, \quad m=0,1,2,3, \quad s=5,6 .
\label{equations}
\end{eqnarray}
The most general solution for a free particle in $d=1+5$ should be written as a superposition
of all  four ($2^{6/2 -1}$) states of a single Weyl representation. We ask the reader to see 
Appendix \ref{appendixtechnique} for the technical details on how to write a Weyl representation 
in terms of the Clifford algebra objects after making a choice of the Cartan subalgebra,  
for which we take: $S^{03}, S^{12}, S^{56}$. 
In our technique \cite{holgernorma2002} one spinor representation---the four 
states, the eigenstates of the chosen 4Cartan subalgebra---are expressed with 
the following four products of projections $\stackrel{ab}{[k]}$ and nilpotents 
$\stackrel{ab}{(k)}$: 
\begin{eqnarray}
\varphi^{1}_{1} &=& \stackrel{56}{(+)} \stackrel{03}{(+i)} \stackrel{12}{(+)}\psi_0,\nonumber\\
\varphi^{1}_{2} &=&\stackrel{56}{(+)}  \stackrel{03}{[-i]} \stackrel{12}{[-]}\psi_0,\nonumber\\
\varphi^{2}_{1} &=&\stackrel{56}{[-]}  \stackrel{03}{[-i]} \stackrel{12}{(+)}\psi_0,\nonumber\\
\varphi^{2}_{2} &=&\stackrel{56}{[-]} \stackrel{03}{(+i)} \stackrel{12}{[-]}\psi_0,
\label{weylrep}
\end{eqnarray}
where  $\psi_0$ is a vacuum state.
If we write 
the operators of handedness in $d=1+5$ as $\Gamma^{(1+5)} = \gamma^0 \gamma^1 
\gamma^2 \gamma^3 \gamma^5 \gamma^6$ ($= 2^3 i S^{03} S^{12} S^{56}$), in $d=1+3$ 
as $\Gamma^{(1+3)}= -i\gamma^0\gamma^1\gamma^2\gamma^3 $ ($= 2^2 i S^{03} S^{12}$) 
and in the two dimensional space as $\Gamma^{(2)} = i\gamma^5 \gamma^6$ 
($= 2 S^{56}$), we find that all four states are lefthanded with respect to 
$\Gamma^{(1+5)}$, with the eigenvalue $-1$, the first two are righthanded and the second two 
lefthanded with respect to 
$\Gamma^{(2)}$, with  the eigenvalues $1$ and $-1$, respectively, while the first two are lefthanded 
and the second two righthanded with respect to $\Gamma^{(1+3)}$ with the eigenvalues $-1$ and $1$, 
respectively. 

Taking into account Eq.(\ref{weylrep}) we may write a wave function  $\psi^{(6)}$ in $d=1+5$ as
\begin{eqnarray}
\psi^{(6)} = (\A \,{\stackrel{56}{(+)}} + \B \,{\stackrel{56}{[-]}})\, \psi^{(4)}, 
\label{psi6}
\end{eqnarray}
where $\A$ and ${\cal B}$ depend on $x^5$ and $x^6$, while $\psi^{(4)}$ determines the spin 
and the coordinate dependent part of the wave function $\psi^{(6)}$ in $d=1+3$. 

Spinors which manifest masslessness in $d=1+3$ must obey the equation
\begin{eqnarray}
\gamma^0\gamma^s p_s \psi^{(6)}=0,\quad s=5,6, 
\label{m}
\end{eqnarray}
since what will demonstrate as an effective action in $d=1+3$ is
\begin{eqnarray}
\int \prod_m dx^m  \Tr\Sb{0123}(\int dx^5 dx^6 \Tr\Sb{56} (\psi^{(6)\dagger}
\gamma^0 (\gamma^m p_m + \gamma^s p_s) \psi^{(6)})) = \nonumber\\
\int \prod_m dx^m \Tr\Sb{0123}(\psi^{(4)\dagger} 
\gamma^0 \gamma^m p_m \psi^{(4)} ) - \int \prod_m dx^m \Tr\Sb{0123}(\psi^{(4)\dagger} 
\gamma^0 m \psi^{(4)} ),
\label{m1}
\end{eqnarray}
where integrals go over all the space on which the solutions are defined. 
$ \Tr\Sb{0123} $ and $ \Tr\Sb{56} $ mean the trace over the spin degrees 
of freedom in $x^0, x^1, x^2, x^3$ and in $x^5,x^6$, respectively. 
(One finds, for example, that $ \Tr\Sb{56} (\stackrel{56}{[\pm]})=1.$)

For massless spinors it must be that  $\int dx^5 dx^6 \Tr_{56} 
(\psi^{(6)\dagger}\gamma^0\gamma^s p_s \psi^{(6)})=
\psi^{(4)\dagger}\gamma^0(-m) \psi^{(4)}$$ = 0$.  
To find the effective action in $1+3$ for massive spinors we recognize that  
$\psi^{(4)\dagger} \gamma^0 (-{\cal A}^* \stackrel{56\;}{(+)^{\dagger}}
+ \,{\cal B}^* \stackrel{56\;}{[-]^{\dagger}})\gamma^s p_s 
({\cal A} \stackrel{56}{(+)}$ $
+ \,{\cal B} \stackrel{56}{[-]})\psi^{(4)}$ $ = 
\psi^{(4)\dagger} \gamma^0 (-{\cal A}^* \stackrel{56\;}{(+)^{\dagger}}
+\, {\cal B}^* \stackrel{56\;}{[-]^{\dagger}})(-m) 
(-{\cal A} \stackrel{56}{(+)}
+\, {\cal B} \stackrel{56}{[-]})$ $\psi^{(4)}$, 
with $s=5,6$, $\stackrel{56\;}{(\pm)^{\dagger}}=
- \stackrel{56}{(\mp)}$ and  $\stackrel{56\;}{[\pm]^{\dagger}}= \stackrel{56}{[\pm]}$, 
while $(^*)$ means complex conjugation. 
We took into account that $\gamma^0 \stackrel{56}{(+)} = - \stackrel{56}{(+)}\gamma^0$, while 
$\gamma^0\stackrel{56}{[-]}= \stackrel{56}{[-]}\gamma^0.$ 
We find that $\Tr_{56}(\stackrel{56\;}{(+)^{\dagger}} \stackrel{56}{(+)})$ = 
$1$. 
In order that 
$\int dx^5 dx^6 \Tr_{56}(\psi^{(4)\dagger} \gamma^0 (-{\cal A}^* \stackrel{56\;}{(+)^{\dagger}}
+ {\cal B}^* \stackrel{56\;}{[-]^{\dagger}})\gamma^s p_s 
({\cal A} \stackrel{56}{(+)}
+ {\cal B} \stackrel{56}{[-]})\psi^{(4)}$ will appear in $d=1+3$ as a mass term $  
\psi^{(4)\dagger} \gamma^0 (-m) \psi^{(4)}$, we must solve the equation 
$\gamma^s p_s \,({\cal A} \stackrel{56}{(+)} + \,{\cal B} \stackrel{56}{[-]}) = 
(-m) (-{\cal A} \stackrel{56}{(+)} + {\cal B} \stackrel{56}{[-]})$.

We can rewrite equations of motion in terms 
of the two complex superposition of $x^5$ and $x^6$, $z: = x^5 + ix^6$ and  $\bar{z}: =  
x^5 - ix^6$
and their derivatives, defined as $\frac{\partial}{\partial z}: = \frac{1}{2}(
\frac{\partial}{\partial x^5} -i \frac{\partial}{\partial x^6}), $ 
$\frac{\partial}{\partial \bar{z}}: = \frac{1}{2}(
\frac{\partial}{\partial x^5} + i \frac{\partial}{\partial x^6}) $ and in terms of the 
two projectors $\stackrel{56}{[\pm]}: = \frac{1}{2}(1\pm i \gamma^5 \gamma^6)$
as follows
\begin{eqnarray}
2i\gamma^5 \{ \frac{\partial}{\partial z} \stackrel{56}{[-]} 
+ \frac{\partial}{\partial \bar{z}} \stackrel{56}{[+]} \}
({\cal A} \stackrel{56}{(+)} + {\cal B} \stackrel{56}{[-]})= 
- m (-{\cal A} \stackrel{56}{(+)} + {\cal B} \stackrel{56}{[-]}).  
\label{equationsin56}
\end{eqnarray}
Since in Eq.(\ref{equationsin56}) $\psi^{(4)}$ would be just  a spectator, we skipped it.

In the massless case the superposition of the first two states 
($\psi^{(6)m=0}_{+}=
\stackrel{56}{(+)} \psi^{(4)m=0}_{+}$, with $\psi^{(4)m=0}_{+} = (\alpha \stackrel{03}{(+i)} 
\stackrel{12}{(+)} + \beta \stackrel{03}{[-i]} \stackrel{12}{[-]})\psi_0$) 
or the second two states ( $\psi^{(6)m=0}_{-}=
\stackrel{56}{[-]} \psi^{(4)m=0}_{-}$, with $\psi^{(4)m=0}_{-} = (\alpha \stackrel{03}{[-i]} 
\stackrel{12}{(+)} + \beta \stackrel{03}{(+i)} \stackrel{12}{[-]})\psi_0$) 
of the left handed Weyl representation presented in Eq.(\ref{weylrep}) must be taken, with the ratio 
of the two parameters $\alpha$ and $\beta$ 
determined by the dynamics in $x^m$ space. In the  massive case  
$\psi^{(6)m}$ is the superposition of 
all the states to which $\gamma^5$ and $\gamma^0$ separately transform the starting state:  
$\psi^{(6)m} = ({\cal A} \stackrel{56}{(+)} + {\cal B} \stackrel{56}{[-]}) \psi^{(4)m}_{\pm}$, 
with $\psi^{(4)m}_{\pm} = \{\alpha [ \stackrel{03}{(+i)} 
\stackrel{12}{(+)} \pm  \stackrel{03}{[-i]} \stackrel{12}{(+)}] + \beta
[\stackrel{03}{[-i]} \stackrel{12}{[-]} \pm \stackrel{03}{(+i)} \stackrel{12}{[-]}]\}\psi_0. $
The sign $\pm$ denotes the eigenvalue of $\gamma^0$ on these states.

We shall therefore simply write (as suggested in Eq.(\ref{psi6}))
$\psi^{(6)} = ({\cal A} \stackrel{56}{(+)} + {\cal B} \stackrel{56}{[-]}) \psi^{(4)}$
in the massless and the massive case, taking into account that in the massless case either 
${\cal A}$ or ${\cal B}$ is nonzero, while in the massive case both are nonzero. Accordingly 
also $\psi^{(4)}$ differs in the massless and the massive case.

We want our states to be eigenstates of the total angular momentum operator $M^{56}$ around 
a chosen origin in the flat 
two dimensional manifold ($M^{(2)}$)  
\begin{eqnarray}
M^{56} = z \frac{\partial}{\partial z} - \bar{z} \frac{\partial}{\partial \bar{z}} + 
S^{56}.
\label{mab}
\end{eqnarray}

Taking into account that $\gamma^5 \stackrel{56}{(+)} = - \stackrel{56}{[-]}, $ 
$\gamma^5 \stackrel{56}{[-]} =  \stackrel{56}{(+)},$  
(see Appendix\ref{appendixtechnique}) 
we end up with the equations for ${\cal A}$ and ${\cal B}$
\begin{eqnarray}
\frac{\partial {\cal B}}{\partial z} + \frac{im}{2} {\cal A} =0,\nonumber\\
\frac{\partial {\cal A}}{\partial \bar{z}}  + \frac{im}{2} {\cal B}=0.  
\label{equationsin56red1}
\end{eqnarray}

For $m=0$ we get as solutions 
\begin{eqnarray}
\psi^{(6)m=0}_{n+1/2} &=& a_n z^n \stackrel{56}{(+)} \psi^{(4)}_{+},\nonumber\\
\psi^{(6)m=0}_{-(n+1/2)} &=& b_n \bar{z}^n \stackrel{56}{[-]} \psi^{(4)}_{-},\; n \ge 0.
\label{solmeq0}
\end{eqnarray}
We required $n\ge 0$ to ensure the integrability of solutions at the origin. The solutions
have  the eigenvalues of $M^{56}$ equal to $(n+1/2)$ and $-(n+1/2)$, respectively. 
Having solutions of both handedness we conclude that there is no mass protection.

For $m\ne0$ we get
\begin{eqnarray}
\psi^{(6)m}_{\pm n+1/2} = a_n (J_n  \stackrel{56}{(+)}   -i J_{n+1}e^{i\phi} 
\stackrel{56}{[-]})e^{\pm in\phi} \psi^{(4)m},\; {\rm for \; n \ge 0},
\label{solmeqm}
\end{eqnarray}
where $J_n$ is the Bessel's functions of the first order. 
The easiest way to see that  $J_n$ and $J_{n+1}$ determine the massive solution  
is to  use Eq.(\ref{equationsin56red1}), take into account that $z=\rho e^{i\phi}$, 
define  $r=m \rho, \rho=\sqrt{(x^5)^2 + (x^6)^2}$,
recognize that $\frac{\partial}{\partial z} = \frac{1}{2} e^{-i\phi} (\frac{\partial}{\partial \rho} 
- \frac{i}{\rho} \frac{\partial}{\partial \phi})$ and we find 
 ${\cal B}= - \frac{2}{im} \frac{\partial {\cal A}}{\partial \bar{z}}$. 
 Then for the choice ${\cal A} = J_n 
e^{in \phi}$ it follows  that ${\cal B} = -i e^{i(n+1)\phi} (\frac{n}{r} J_{n} -  
 \frac{\partial J_n}{\partial r}) $, which tells that  
${\cal B} = -i J_{n+1} e^{i(n+1)\phi}.$

\section{Boundary conditions and involution}
\label{involution}

In the ref.\ \cite{hnkk06} we make a choice of particular solutions of the equations of motion 
by requiring that  $\hat{\cal{R}}\psi|_{\rm wall} = 0,$ Eq.(\ref{diskboundary}),
where the wall was put on the circle 
of the radius $\rho_0$ of 
the finite disk (Eqs.(\ref{diskboundary},\ref{prodiskboundary})). 
This boundary condition requires that 
in the massless case (since $\stackrel{56}{[-]} \stackrel{56}{(+)} =0$ while 
$\stackrel{56}{[-]} \stackrel{56}{[-]} = \stackrel{56}{[-]} $)  only the right handed 
solutions (Eq.\ref{solmeq0}) $\psi^{(6)m=0}_{n+1/2} = a_n z^n \stackrel{56}{(+)} \psi^{(4)m=0}_{+}$ 
(that is the left handed with respect to $SO(1,3)$) are allowed, while the left 
handed solutions must be zero ($b_n=0$) making the mass protection mechanism work in $d=1+3$.  
This boundary condition allows all the angular momenta $M^{56} = 1/2,3/2,\cdots,$ which is 
still not what we would expect to have as a charge of massless spinors in 
$1+3$, namely $n=1/2$ only. 

In the massive case the boundary condition determines masses of solutions, since only the 
solutions with $J_{n+1}|_{\rho = \rho_0} =0$ are allowed from the same reason as discussed 
for the massless case. This boundary condition determines masses of spinors through the relation 
$m_{i n+1/2} \rho_0$ is equal to a $i^{th}$ zero of $J_{n+1}$ 
($J_{n+1}(m_{i n+1/2} \rho_0)=0$). In the massive 
case all the zeros of any $J_{n+1}(m_{i n+1/2} \rho_0)=0$ contribute.

This time we look for the {\em involution boundary conditions}.
First we recognize that for a flat $M^2$ 
the $Z_2$ or involution symmetry can be recognized: {\em The transformation 
$\rho/\rho_{0} \rightarrow \frac{\rho_0}{\rho}$} (which can be written also as $z/\rho_0 \rightarrow 
\frac{\rho_0}{\bar{z}}$) {\em transforms the exterior of the disk into the interior of the disk} 
and conversely. 

Then we extend the involution operator to operate also on the space of solutions
\begin{eqnarray}
\hat{{\cal O}} &=& (I - 2 \hat{{\cal R'}})|_{z/\rho_0 \rightarrow 
\rho_0/\bar{z}}, \nonumber\\
\hat{{\cal O}}^2 &=& I.
\label{involutionO}
\end{eqnarray}
The involution condition $\hat{{\cal O}}^2 =I$ requires, that $\hat{{\cal R'}}$ 
is a projector 
\begin{eqnarray}
(\hat{{\cal R'}})^2=  \hat{{\cal R'}}
\label{involutionR}
\end{eqnarray}
and can be written as $\hat{{\cal R'}}=\hat{{\cal R}} +  \hat{{\cal R}}_{add}$, 
where $\hat{{\cal R}}_{add}$ must be a nilpotent operator  fulfilling 
the conditions 
\begin{eqnarray}
(\hat{{\cal R}}_{add})^2= 0,\;\; \hat{{\cal R}}_{add} \hat{{\cal R}} =0,\;\;
\hat{{\cal R}} \hat{{\cal R}}_{add} = \hat{{\cal R}}_{add}, 
\label{involutionRadd}
\end{eqnarray}
 We had $\hat{{\cal R}} = \stackrel{56}{[-]},$ which is the projector. 
Since we find that $ \stackrel{56}{[-]} \stackrel{56}{(-)} = \stackrel{56}{(-)}$ 
(see Appendix \ref{appendixtechnique}), while 
$ \stackrel{56}{(-)} \stackrel{56}{[-]} = 0,$ we can choose $\hat{{\cal R}}_{add} = \alpha
 \stackrel{56}{(-)}$, where $\alpha$ is any function of  $z$ and $\frac{\partial}{\partial z}$. 
Let us point out that $\hat{{\cal R}}_{add}$ is not a Hermitean operator, since 
$\stackrel{56\;}{(-)^{\dagger}}= -\stackrel{56}{(+)}$ and $z^{\dagger} = \bar{z}, 
(\frac{\partial}{\partial z})^{\dagger}
=\frac{\partial}{\partial {\bar z}}$. Accordingly also neither $\hat{{\cal R'}}$ nor 
$\hat{{\cal O}}$ is a Hermitean operator.

We now make a choice of a natural boundary conditions on the wall $\rho = \rho_0$ 
\begin{eqnarray}
\{\hat{{\cal O}} \psi&=& \psi \}|_{\rm wall}, 
\label{involutionOwall}
\end{eqnarray}
saying that what ever the involution operator is, the state $\psi $ and its 
involution $\hat{{\cal O}}\psi$ must be the same on the wall, that is at $\rho=\rho_0$.

It is worthwhile to write the involution operator $\hat{{\cal O}}$ and correspondingly the 
projector $\hat{{\cal R'}}$ in a covariant way. 
Recognizing that $n^{(\rho)}{}_{\!\! a} \,\gamma^a n^{(\rho)}{}_{\!\! b} \, p^b =$  
$i\{(e^{2i\phi} \frac{\partial}{\partial z} + \frac{\partial}{\partial \bar{z}}) 
\stackrel{56}{(-)} +  (\frac{\partial}{\partial z} + e^{-2i\phi} 
\frac{\partial}{\partial \bar{z}})  \stackrel{56}{(+)}\} $
we may write 
\begin{eqnarray}
\label{Rprime}
\hat{\cal R}' &=& \frac{1}{2} (1-i n^{(\rho)}{\!\!}_{a} \,n^{(\phi)}{\!\!}_{b} \,
\gamma^a \gamma^b ) (1 - \beta n^{(\rho)}{\!\!}_a \,\gamma^a \,n^{(\rho)}{\!\!}_b \,p^b )\nonumber\\
&=& 
\stackrel{56}{[-]}( I - \beta i e^{i\phi} \frac{\partial}{\partial \rho}\stackrel{56}{(-)}).
\end{eqnarray}
 This is just 
our generalized projector $\hat{{\cal R'}}$, if we  make a choice for   
$\alpha$ from Eq.(\ref{involutionRadd}) as follows: $\alpha = -\beta i 
(e^{2i\phi} \frac{\partial}{\partial z} +  
\frac{\partial}{\partial \bar{z}})$
(since $\stackrel{56}{[-]}\stackrel{56}{(-)} =
\stackrel{56}{(-)})$, while $\stackrel{56}{[-]}\stackrel{56}{(+)} = 0$),  
where $\beta $ is any complex number. 

The projector $\hat{{\cal R'}} $ (Eq.(\ref{Rprime})) entering into the orbifolding condition
 $( \hat{\cal O}\psi)|_{wall}= (\psi)|_{wall}$,
with $ \hat{\cal O} =I- 2\hat{\cal R}' $, requires that
 \begin{eqnarray}
 \label{Rprime1}
(\stackrel{56}{[-]}\{I-i \beta e^{i\phi } \frac{\partial}{\partial \rho} \stackrel{56}{(-)} \} 
\psi)|_{wall}=0.
\end{eqnarray}

For the massless case we obtain:
$(\stackrel{56}{[-]}[I-i\beta e^{i\phi} 
\frac{\partial}{\partial \rho} \stackrel{56}{(-)}]z^n \stackrel{56}{(+)})|_{wall}$ $
= $ $(-i\beta e^{2i\phi} n z^{n-1}\stackrel{56}{[-]})|_{wall}=0$, 
{\em which has the solution for an arbitrarily chosen $\beta$ only if
$n=0.$}

For the massive case we have:
$(\stackrel{56}{[-]}\{I-i \beta e^{i\phi } \frac{\partial}{\partial \rho} \stackrel{56}{(-)} \}
(J_n \stackrel{56}{(+)}-i  J_{n+1} e^{i\phi}\stackrel{56}{[-]}) e^{in\phi})|_{wall}=
((-i J_{n+1} +i \beta  \frac{\partial J_n}{\partial \rho})e^{i(n+1)\phi } \stackrel{56}{[-]})|_{wall}=0,$
which has again {\em the solution for an arbitrarily chosen $\beta$ only for $n=0$}.  
In this case 
we namely have
$J_1= - \frac{\partial J_{0}}{\partial m_{i m+1/2}\rho}$.
If we chose that $J_1|_{wall}=0, $  then 
$\frac{\partial J_{0}}{\partial m_{i n+1/2}\rho}|_{wall} =0$ and
the relation
$(-J_{1} + \beta  \frac{\partial J_0}{\partial \rho})|_{wall}= 0$ for any $\beta$.
This relation is fulfilled for infinite many masses $m_{i 1/2}, i=1,\cdots$, where index $i$ 
counts the  zeros of $J_1$ determined by the relation
 $J_1(m_{i1/2}\rho)|_{wall}=0. $

We conclude this section by recognizing that  we have on the disk 
{\em only one massless solution}, namely 
\begin{eqnarray}
\label{onemassless}
\psi^{m=0}_{1/2}= a_0 \stackrel{56}{(+)}
\end{eqnarray}
{\em infinite many massive solutions}, namely
\begin{eqnarray}
\label{massive}
\psi^{m^i}_{1/2}= a_i (J_{oi}(\alpha_{1i} \rho/\rho_0) \stackrel{56}{(+)} - iJ_{1i}
(\alpha_{1i} \rho/\rho_0)),
\end{eqnarray}
with $i$ which denotes the $i$-th zero of $J_{\alpha_{1i}}=0$. All the solutions have 
the eigenvalue of $M^{56}$ equal to $1/2$ and obey 
the equations of motion (Eq.(\ref{equationsin56red1})) {\em and the orbifolding 
boundary condition} (Eq.(\ref{involutionOwall})) with $\hat{\cal R}'$  from Eq.(\ref{Rprime}). 
We get the corresponding solutions in $d=1+5$ ($\psi^{(6)}$) by multiplying the wave functions 
of Eqs.(\ref{onemassless},\ref{massive}) with the corresponding $\psi^{(4)}$, which distinguish  
among the massless and the massive solutions as described in Sect.\ref{equations}.

\section{Current through the wall}
\label{current}

The current perpendicular to the wall can be written as
\begin{eqnarray}
n^{(\rho)s} j_s &=&\psi^{\dagger} \gamma^0 \gamma^s n^{(\rho)}_s \psi =  
\psi^{\dagger}\hat{j}_{\perp} \psi, 
\quad \hat{j}_{\perp} = - \gamma^0 \{ e^{-i\phi} \stackrel{56}{(+)} + 
e^{i\phi} \stackrel{56}{(-)}\}. 
\label{current}
\end{eqnarray}
We need to know the current through the wall, which for physically acceptable cases when 
spinors are localized inside the disk (involution transforms outside the 
disk into its inside, or equivalently, it transforms inside the 
disk into its outside)  must be zero.
We accordingly expect that the current through the wall is equal to zero:
\begin{eqnarray}
\{\psi^{\dagger}\hat{j}_{\perp}\psi\}|_{\rm wall} = 0 = 
 \{\psi^{\dagger} {\hat{\cal O}}^{\dagger}\hat{j}_{\perp}{\hat{\cal O}} \psi\}|_{\rm wall}. 
\label{current}
\end{eqnarray}

Since $\hat{{\cal O}}^{\dagger} = I - 2 (\hat{{\cal R}} + \hat{{\cal R}}^{\dagger}_{add})$
and $\hat{{\cal R}}^{\dagger}_{add} = (-i \beta e^{i\phi } 
\frac{\partial}{\partial \rho} \stackrel{56}{(-)})^{\dagger} = 
-i \beta^* e^{-i\phi } \frac{\partial}{\partial \rho} \stackrel{56}{(+)}$, it follows that
$\hat{{\cal O}}^{\dagger}\hat{j}_{\perp}\hat{{\cal O}} = - \hat{j}_{\perp}
- 2i\gamma^0 \frac{\partial}{\partial \rho} ( \beta^* \stackrel{56}{[-]} - 
\beta  \stackrel{56}{[+]})$.

It must then be
\begin{eqnarray}
\{\psi^{\dagger}\hat{j}_{\perp}\psi\}|_{\rm wall} =  0 = 
 -( \psi^{\dagger} \{\hat{j}_{\perp}  + 2i \gamma^0 \frac{\partial}{\partial \rho}
 (\beta^* \stackrel{56}{[-]} - 
\beta  \stackrel{56}{[+]}) \} \psi )|_{\rm wall} 
\label{currentwall}
\end{eqnarray}
for any $\beta $ and any superposition of the states obeying our involution boundary 
condition of Eq.(\ref{involutionOwall}), with $\hat{{\cal O}}$ (Eq.\ref{involutionO}) 
and $\hat{{\cal R'}}$ (Eq.\ref{Rprime}). 
One easily finds that the current through the wall (Eq.(\ref{currentwall})) is equal 
to zero for the massless and all the massive solutions (or any superposition of solutions) 
obeying the involution 
boundary condition (Eq.(\ref{involutionOwall})).

\section{Hermiticity of the operators and orthogonality of solutions}
\label{hermiticity}

In this section we redefine the operators $ p_s$ and $(\gamma^s p_s)^2$ so that they become 
Hermitean on the vector space of the states fulfilling 
 our particular involution boundary conditions (Eqs.(\ref{involutionO},\ref{Rprime})),  
that is on the vector space of the massless state $a_0 \stackrel{56}{(+)}$ 
(Eq.\ref{onemassless}) and  the  massive 
states $a_{i} (J_{0i} \stackrel{56}{(+)} -iJ_{1i}\stackrel{56}{[-]}e^{i\phi})$ (Eq.(\ref{massive})), 
where $i$ counts the $i-$th zero  $\alpha_{1i}$ of  $J_{1i}$. All the states have the eigenvalue 
of $M^{56}$ equal to $1/2$. We also comment 
on the orthogonality properties of  these states. 

 We  expect that on the space of all the states fulfilling our  involution boundary conditions \\   
i) the operators $ p_s$ are Hermitean,\\
 ii) the states are accordingly orthogonal 
 ($ \int d^2x 
 \psi_{i}{\!}^{\dagger}(\gamma^s p_s)^2)  \psi_j = 
\int d^2x \psi_{i}{\!}^{\dagger}  \psi_j m^2  \delta_{ij}$).

First we recognize that the operator $p_s$ is not Hermitean  on the states, which are 
not zero on the wall (that is at $\rho=\rho_0$) since for those state  
$\int d^2x \psi_{i}{\!}^{\dagger} p_s \psi_{j} +  
 \int d^2x (-p_s \psi_{i})^{\dagger} \psi_{j}\ne 0$. 
 
 We therefore replace $p_s$ with $\hat{p}_s$ 
   \begin{eqnarray}
   \hat{p}_s= i\{ \frac{\partial}{\partial x^s} -  \frac{1}{2} \frac{x^s}{\rho}
   \delta(\rho-\rho_0)\stackrel{56}{[+]}\}. 
   \label{ps}
   \end{eqnarray}
One finds, for example, that $\Tr\Sb{56} \int d^2x \psi_{j}^{\dagger} (\hat{p}_s
\psi_i)= \Tr\Sb{56} \int d^2x (\hat{p}_s\psi_j)^{\dagger} 
\psi_i = -i\pi/2 \rho^{2(n+1)}_{0} \varepsilon$, with $\varepsilon =1, -i$ for $s=5,6,$ 
respectively when  
$\psi_i= \rho^n e^{in \phi} \stackrel{56}{(+)}$ and $\psi_j= \rho^{n+1} e^{i(n+1)\phi}
\stackrel{56}{(+)} $.  

Since $\stackrel{56}{(+)}$ (the massless state) and $\stackrel{56}{[-]}$ are orthogonal 
in the spin part, while the matrix element  of  $ \hat{p}_s$ is zero between the 
massless state  
and $J_{0i} \stackrel{56}{(+)}$,  
we check instead the hermiticity properties of the operator $(\gamma^s \hat{p}_s)^2$  
  \begin{eqnarray}
  \gamma^s \hat{p}_s \gamma^t \hat{p}_t &=& p_s p^s \nonumber\\
  &+&\frac{1}{2}\{ [\frac{\partial}{\partial \rho}\delta(\rho -\rho_0) + \frac{1}{\rho} 
  \delta(\rho-\rho_0) + \delta(\rho-\rho_0)
  ( \frac{\partial}{\partial \rho} - \frac{i}{\rho} \frac{\partial}{\partial \phi})]
  \stackrel{56}{[+]}\nonumber\\ 
  &+& \delta(\rho-\rho_0) (\frac{\partial}{\partial \rho} -  
 \frac{i}{\rho}  \frac{\partial}{\partial \phi})\stackrel{56}{[-]}\},
  \label{pssquared}
  \end{eqnarray}
where $p_s p^s= \frac{\partial^2}{\partial \rho^2} + \frac{1}{\rho^2}
\frac{\partial^2}{\partial \phi^2} +  \frac{1}{\rho}
\frac{\partial}{\partial \phi}$.

Let us first check the Hermiticity and accordingly the orthogonality relations 
between the massless and  any of the massive states obeying our involution boundary condition. 
Since the spinor parts $\stackrel{56}{(+)}$ and $\stackrel{56}{[-]}$ are orthogonal,
we only have to check the Hermiticity properties with respect to 
the $\stackrel{56}{(+)}$ component of the massive states. We find that 
\begin{eqnarray}
\label{herorth0}
\int d^2x \Tr\Sb{56} (J_{0i}\stackrel{56}{(+)})^{\dagger} 
[(\gamma^s \hat{p}_s)^2 
\stackrel{56}{(+)}] =  \pi \frac{\alpha_{1i}}{\rho_0} (\rho J_{1i})|_{\rho=\rho_0} = 0 \nonumber\\
= \int d^2x \Tr\Sb{56} [(\gamma^s \hat{p}_s )^2 J_{0i}\stackrel{56}{(+)}]^{\dagger}  
\stackrel{56}{(+)} = \pi (-\frac{\alpha_{1i}}{\rho_0} \rho J_{1i})|_{\rho=\rho_0},
\end{eqnarray}
due to the properties 
of the Bessel functions $J_{1i}= - \frac{\rho_0}{\alpha_{1i}} 
\frac{\partial J_{0i}}{\partial \rho}$ with 
$J_{1i}(\alpha_{1i})= 0$ (due to our particular boundary condition), 
and of the delta function $\int_{0}^{\rho_0} f(\rho) 
\frac{\partial \delta(\rho -\rho_0)}{\partial \rho} = - \frac{\partial f}{\partial \rho}$, 
while $m_{i}=\frac{\alpha_{1i}}{\rho_0}$. 
Accordingly the massless state 
is orthogonal to all the massive states. 

We also find that the operator $(\gamma^s \hat{p}_s )^2 $ is Hermitean on the space of massive states. 
Taking into account that for the  Bessel functions 
$\frac{\partial J_{1i}}{\partial \rho} = \frac{\alpha_{1i}}{\rho_0}
J_{0i} + \frac{\rho_0}{\alpha_{1i}} \frac{1}{\rho} \frac{\partial J_{0i}}{\partial \rho}$
one finds that for $i\ne k$ it follows  
\begin{eqnarray}
\label{herorth1}
&&\int d^2x \Tr\Sb{56} 
(J_{0i}\stackrel{56}{(+)} -iJ_{1i} \stackrel{56}{[-]} e^{i \phi})^{\dagger} 
[(\gamma^s \hat{p}_s)^2 (J_{0k}\stackrel{56}{(+)}-i J_{1k} \stackrel{56}{[-]} e^{i \phi})]=\nonumber\\ 
&&2 \pi \{\int^{\rho_0}_0 \rho d\rho [- (m_{k})^2 (J_{0i}J_{0k} + J_{1i}J_{1k})] 
+ \; \frac{1}{2} ( -\rho \frac{\partial J_{0i}}{\partial \rho} J_{0k} + \rho J_{1i} J_{0k} 
\frac{\alpha_{1k}}{\rho_0})|_{\rho = \rho_0} \}= \nonumber\\
&&2\pi(\rho J_{0i}J_{1k} + 
\rho J_{1i}J_{0k})|_{\rho=\rho_0} =0, 
\end{eqnarray} 
 since $ J_{1k} (\alpha_{1k})=0.$ We checked accordingly 
the Hermiticity of the operator $(\gamma^s \hat{p}_s )^2 $ on the vector space of 
the massive states and correspondingly  also the orthogonality of these states.

Let us add  that 
\begin{eqnarray}
\label{orth0}
&&\int d^2x \Tr\Sb{56} 
(J_{0i}\stackrel{56}{(+)} -iJ_{1i} \stackrel{56}{[-]} e^{i \phi})^{\dagger} 
[(\gamma^s \hat{p}_s ]^2 (J_{0i}\stackrel{56}{(+)}-i J_{1i} \stackrel{56}{[-]} e^{i \phi})=\nonumber\\
&&- (m_{k})^2 \pi (\rho^2 (J_{0i}^2  +  J_{1i}^2))|_{\rho = \rho_0} = 
\pi (\rho^2 J_{0i}^2)|_{\rho=\rho_0},\nonumber\\
&&\int d^2x \Tr\Sb{56} 
(\stackrel{56}{(+)})^{\dagger} 
[(\gamma^s \hat{p}_s ]^2 \stackrel{56}{(+)}=0. 
\end{eqnarray}

So we conclude that on the vector space of all the states, which obey our 
particular boundary condition (Eqs.(\ref{involutionO},\ref{Rprime})) 
the operator $(\gamma^s \hat{p}_s)^2$ (Eqs.(\ref{ps},\ref{pssquared})) is Hermitean and the 
states obeying also the equations of motion (Eqs.(\ref{onemassless},\ref{massive}),  are orthogonal.

Looking at Eq.(\ref{orth0}),  we recognize, that $m_{i}= \frac{\alpha_0i}{\rho_0}$ determine the 
masses of the states $\psi^{(6)m}_{1/2 i}$. We read in Eq.(\ref{orth0}) the normalization 
factor of the massive  states, while the massless state has to be normalized according to 
the relation $\int d^2x \Tr\Sb{56} 
(\stackrel{56}{(+)})^{\dagger} \stackrel{56}{(+)}= 2 \pi \frac{\rho_{0}^2}{2}$.

\section{Properties of spinors in $d=1+3$}
\label{properties1+3}

To study how do spinors couple to the Kaluza-Klein gauge fields in the case of $M^{(1+5)}$, ``broken'' to 
$M^{(1+3)} \times $ a flat disk with $\rho_0$ and with the involution boundary condition, 
which allows only right handed spinors
at $\rho_0$,
we first look for (background) gauge gravitational fields, which preserve the rotational symmetry 
on the disk. Following ref.\ \cite{hnkk06} we find 
for the background vielbein field  
\begin{eqnarray}
e^a{}_{\alpha} = 
\pmatrix{\delta^{m}{}_{\mu}  & e^{m}{}_{\sigma}=0 \cr
 e^{s}{}_{\mu} & e^s{}_{\sigma} \cr},
f^{\alpha}{}_{a} =
\pmatrix{\delta^{\mu}{}_{m}  & f^{\sigma}{}_{m} \cr
0= f^{\mu}{}_{s} & f^{\sigma}{}_{s} \cr},
\label{f6}
\end{eqnarray}
with $f^{\sigma}{}_{m} = A_{\mu} \delta ^{\mu}{}_{m}
\varepsilon^{\sigma}{}_{\tau} x^{\tau}$
and  the spin connection field 
\begin{eqnarray}
\omega_{st \mu} = - \varepsilon_{st}  A_{\mu},\quad \omega_{sm \mu} = 
-\frac{1}{2} F_{\mu \nu} \delta^{\nu}{}_{m}
\varepsilon_{s \sigma} x^{\sigma}.
\label{omega6}
\end{eqnarray}
 The $U(1)$ gauge field $A_{\mu}$ depends only on $x^{\mu}$.
All the other components of the spin connection fields are zero, since for simplicity we allow no gravity in
$(1+3)$ dimensional space.

To determine the current, coupled to the Kaluza-Klein gauge fields $A_{\mu}$, we
analyze the spinor action
\begin{eqnarray}
{\cal S} &=& \int \; d^dx E \bar{\psi}^{(6)} \gamma^a p_{0a} \psi^{(6)} = \int \; 
d^dx  \bar{\psi}^{(6)} \gamma^m \delta^{\mu}{}_{m} p_{\mu} \psi^{(6)} + \nonumber\\
&& \int \; d^dx   \bar{\psi}^{(6)} \gamma^m (-)S^{sm} \omega_{sm \mu} \psi^{(6)}  + 
\int \; d^dx  \bar{\psi}^{(6)} \gamma^s \delta^{\sigma}{}_{s} p_{\sigma} \psi^{(6)} +\nonumber\\
&& \int \; d^dx   \bar{\psi}^{(6)} \gamma^m  \delta^{\mu}{}_{m} A_{\mu} 
(\varepsilon^{\sigma}{}_{\tau} x^{\tau}
 p_{\sigma} + S^{56}) \psi^{(6)}.
\label{spinoractioncurrent}
\end{eqnarray}
 $\psi^{(6)}$ are solutions of the Weyl equation in $d=1+3$ .
 $E$ is for $f^{\alpha}{}_{a}$ from (\ref{f6}) equal to 1. 
The first term on the right hand side  of Eq.(\ref{spinoractioncurrent}) is the kinetic term
(together with the last  term defines  
the  covariant derivative $p_{0 \mu}$ in $d=1+3$).  
The second term on the right hand side  contributes nothing when integration over 
the disk is performed, since it is proportional to $x^{\sigma}$ ($\omega_{sm \mu} = -\frac{1}{2}
F_{\mu \nu} \delta^{\nu}{}_{m} \varepsilon_{s \sigma} x^{\sigma}$).

We end up with 
\begin{eqnarray}
j^{\mu} = \int \; d^2x \bar{\psi}^{(6)} \gamma^m \delta^{\mu}{}_{m} M^{56}  \psi^{(6)}
\label{currentdisk}
\end{eqnarray}
as  the current in $d=1+3$.  The charge in $d=1+3$ is  proportional to the total 
angular momentum  $M^{56} =L^{56} + S^{56}$ on a disk, for either massless or massive spinors.


\section{Conclusions}
\label{discussions}

In this paper we were looking for what we call a "natural boundary condition"---a  
condition which would, due to some symmetry relations, make massless spinors which live in  
$M^{1+5}$ and carry nothing but the charge to live in $M^{(1+3)} \times $ a 
flat disk, manifesting in $M^{(1+3)}$, if massless,  
as a left handed spinor (with no right handed partner) and would accordingly be mass 
protected. The spin in $x^5$ and $x^6$ of the left handed massless spinor should  in $M^{(1+3)}$
manifest as the charge  and   
should chirally couple with  the Kaluza-Klein type of charge of only one 
value to the corresponding gauge field. The last requirement ensures that  after 
the second quantization procedure a particle and an antiparticle would not appear each of $\pm $ 
 of the particular charge.

We found the involution boundary condition
\begin{eqnarray}
\{\hat{{\cal O}} \psi&=& \psi \}|_{\rm wall}, \quad 
{\cal {\hat O}} = I-2 \hat{{\cal R'}}, \nonumber\\
\hat{{\cal R'}} &=& \stackrel{56}{[-]}( I + \beta i [e^{2i\phi} \frac{\partial}{\partial z} +  
\frac{\partial}{\partial \bar{z}}]\stackrel{56}{(-)}),
\label{involutionOwallc}
\end{eqnarray}
where $\beta $ is any complex number and can be  
written in a covariant way  as
\begin{eqnarray}
\hat{{\cal R'}} = \frac{1}{2}
 (1-i n^{(\rho)}{\!\!}_{a} \,n^{(\phi)}{\!\!}_{b} \,
\gamma^a \gamma^b ) (1 - \beta n^{(\rho)}{\!\!}_a \,\gamma^a \,n^{(\rho)}{\!\!}_b \,p^b ) = 
\stackrel{56}{[-]}( I + \beta i [e^{2i\phi} \frac{\partial}{\partial z} +  
\frac{\partial}{\partial \bar{z}}]\stackrel{56}{(-)}).
\label{genc}
\end{eqnarray}
$\hat{{\cal O}}$ transforms  solutions of the Weyl equations inside the flat disk 
into outside of it (or conversely) and allows in the massless case only the right  
handed spinor to live  
on the disk and accordingly manifests left handedness in $M^{(1+3)}$.  
The massless and the massive solutions carry in the fifth and sixth dimension  (only) the spin $1/2$, 
which then manifests  as the charge in $d=1+3$. The massless solution is mass protected. 

We defined a generalized momentum $p_s$
  \begin{eqnarray}
  \hat{p}_s=  i\{ \frac{\partial}{\partial x^s} -  \frac{1}{2}\pmatrix{\cos{\phi}\cr
  \sin{\phi} \cr} \delta(\rho-\rho_0) \stackrel{56}{[+]}\},
  \label{psc}
  \end{eqnarray}
which is  Hermitean on the vector space of states obeying our involution boundary condition 
(Eqs.(\ref{involutionOwallc},\ref{involutionO},\ref{Rprime})).

The operator 
$\gamma^s \hat{p}_s \gamma^t \hat{p}_t$ is Heritean
on the vector space of states obeying our boundary condition and 
the solutions of the equations of motion---the massless one (Eq.(\ref{onemassless}) 
and the massive ones (Eq.(\ref{massive})---are accordingly orthogonal, with the eigen 
values of this operator which demonstrate 
the masses of states.

The negative  $-1/2$ charge states appear only after the second quantization 
procedure in agreement with what we observe.

\section{Acknowledgement } One of the authors (N.S.M.B.) would like to warmly thank Jo\v ze Vrabec 
for his very fruitful discussions, which  help a lot to clarify the concept of involution and 
 to use it in the right way.

\appendix %

\section{Spinor representation technique in terms of Clifford algebra objects}
\label{appendixtechnique}

We define\cite{holgernorma2002} spinor representations as superposition of 
products of the Clifford algebra objects 
$\gamma^a$ so that they are  
eigen states of the chosen Cartan sub algebra of the Lorentz algebra $SO(d)$, 
determined by the generators 
$S^{ab} = i/4 (\gamma^a \gamma^b - \gamma^b \gamma^a)$.
By introducing the notation
\begin{eqnarray}
\stackrel{ab}{(\pm i)}: &=& \frac{1}{2}(\gamma^a \mp  \gamma^b),  \quad 
\stackrel{ab}{[\pm i]}: = \frac{1}{2}(1 \pm \gamma^a \gamma^b), \;{\rm  for} \; \eta^{aa} \eta^{bb} =-1, \nonumber\\
\stackrel{ab}{(\pm )}: &= &\frac{1}{2}(\gamma^a \pm i \gamma^b),  \quad 
\stackrel{ab}{[\pm ]}: = \frac{1}{2}(1 \pm i\gamma^a \gamma^b), \;{\rm for} \; \eta^{aa} \eta^{bb} =1,
\label{eigensab}
\end{eqnarray}
it can be checked that  the above binomials are really ``eigenvectors''  of  the generators 
$S^{ab}$
\begin{eqnarray}
S^{ab} \stackrel{ab}{(k)}: &=&  \frac{k}{2} \stackrel{ab}{(k)}, \quad 
S^{ab} \stackrel{ab}{[k]}:  =  \frac{k}{2} \stackrel{ab}{[k]}.
\label{eigensabev}
\end{eqnarray}
Accordingly we have
\begin{eqnarray}
\stackrel{03}{(\pm i)}: &=& \frac{1}{2}(\gamma^0 \mp  \gamma^3),  \quad 
\stackrel{03}{[\pm i]}: = \frac{1}{2}(1 \pm \gamma^0 \gamma^3), \nonumber\\
\stackrel{12}{(\pm )}: &= &\frac{1}{2}(\gamma^1 \pm i \gamma^2),  \quad 
\stackrel{12}{[\pm ]}: = \frac{1}{2}(1 \pm i\gamma^1 \gamma^2), \nonumber\\
\stackrel{56}{(\pm )}: &= &\frac{1}{2}(\gamma^5 \pm i \gamma^6),  \quad 
\stackrel{56}{[\pm ]}: = \frac{1}{2}(1 \pm i\gamma^5 \gamma^6), \nonumber\\
\label{eigensab031256}
\end{eqnarray}
with eigenvalues of $S^{03}$ equal to $\pm \frac{i}{2}$ for $\stackrel{03}{(\pm i)}$ and 
$\stackrel{03}{[\pm i]}$, and to $\pm \frac{1}{2}$ for  $\stackrel{12}{(\pm )}$ and 
$\stackrel{12}{[\pm ]}$, as well as for for $\stackrel{56}{(\pm )}$ and 
$\stackrel{56}{[\pm ]}$. 

We further find 
\begin{eqnarray}
\gamma^a \stackrel{ab}{(k)}&=&\eta^{aa}\stackrel{ab}{[-k]},\quad 
\gamma^b \stackrel{ab}{(k)}= -ik \stackrel{ab}{[-k]}, \nonumber\\
\gamma^a \stackrel{ab}{[k]}&=& \stackrel{ab}{(-k)},\quad \quad \quad
\gamma^b \stackrel{ab}{[k]}= -ik \eta^{aa} \stackrel{ab}{(-k)}.
\label{graphgammaaction}
\end{eqnarray}

We also find 
\begin{eqnarray}
\stackrel{ab}{(k)}\stackrel{ab}{(k)}= 0, & & \stackrel{ab}{(k)}\stackrel{ab}{(-k)}
= \eta^{aa}  \stackrel{ab}{[k]}, \quad 
\stackrel{ab}{[k]}\stackrel{ab}{[k]} =  \stackrel{ab}{[k]}, \;\;\quad \quad
\stackrel{ab}{[k]}\stackrel{ab}{[-k]}= 0, 
 \nonumber\\
\stackrel{ab}{(k)}\stackrel{ab}{[k]} = 0, & &  \stackrel{ab}{[k]}\stackrel{ab}{(k)}
=  \stackrel{ab}{(k)}, \quad \quad \quad
\stackrel{ab}{(k)}\stackrel{ab}{[-k]} =  \stackrel{ab}{(k)},
\quad \quad \stackrel{ab}{[k]}\stackrel{ab}{(-k)} =0.
\label{graphbinoms}
\end{eqnarray}

To represent one Weyl spinor in $d=1+5$, one must make a choice of the
operators belonging to the Cartan sub algebra of $3$ elements of the group $SO(1,5)$ 
\begin{eqnarray}
S^{03}, S^{12}, S^{56}.
\label{cartan}
\end{eqnarray}
Any eigenstate of the Cartan sub algebra (Eq.(\ref{cartan})) must be a product of 
three binomials, each of which is an eigenstate of one of the three elements. 
A left handed spinor ($\Gamma^{(1+5)} = -1$) representation with $2^{6/2-1}$ basic states 
is presented in Eq.(\ref{weylrep}). 
for example, the state  $\stackrel{03}{(+ i)}\stackrel{12}{(+)}
\stackrel{56}{(+)}\psi_0,$ where $\psi_0$ is a vacuum state (any, which is not annihilated 
by the operator in front of the state) has the eigenvalues  of $ S^{03}, S^{12} $ and $S^{56}$ equal 
to $\frac{i}{2}$, $ \frac{1}{2}$ and $\frac{1}{2}$, correspondingly. All the other states
of one representation of $SO(1,5)$ follow from this one by just the application of all possible 
$S^(ab)$, which do not belong to Cartan sub algebra.

\end{document}